\documentclass[12pt]{article}
\usepackage{epsfig,amsfonts,amssymb}
\usepackage{hyperref}
\pdfoutput=1

\usepackage{cite}
\usepackage{cite}

\usepackage{comment}

\def\ZZZ{{\hbox{ Z\kern-1.6mm Z}}}
\def\RRR{{\hbox{ R\kern-2.4mm R}}}
\def\CCC{{\hbox{ C\kern-2.0mm C}}}
\def\zzz{{\hbox{z\kern-1mm z}}}

\newcommand{\nn}{\nonumber \\ }

\newcommand{\qeq}{{\hbox{=\kern-2.3mm ? \kern.5mm }}}
\renewcommand{\qeq}{=}

\newcommand{\eps}{\epsilon}
\newcommand{\vareps}{\varepsilon}

\newcommand{\vp}{\varphi}
\newcommand{\ve}{\varepsilon}

\newcommand{\VV}{{\cal V}}

\newcommand{\AAA}{{\cal A}}

\newcommand{\OO}{{\cal O}}

\newcommand{\LL}{{\cal L}}

\newcommand{\half}{{1\over 2}}
\newcommand{\wt}{\widetilde}

\newcommand{\NN}{{\cal N}}

\newcommand{\SSS}{{\cal S}}

\newcommand{\be}{\begin{equation}}
\newcommand{\ee}{\end{equation}}
\newcommand{\ben}{\begin{eqnarray}\displaystyle}
\newcommand{\een}{\end{eqnarray}}

\newcommand{\refb}[1]{(\ref{#1})}
\newcommand{\p}{\partial}
\newcommand{\sectiono}[1]{\section{#1}\setcounter{equation}{0}}

\def\one{{\hbox{ 1\kern-.8mm l}}}
\def\zero{{\hbox{ 0\kern-1.5mm 0}}}

\newcommand{\bea}[1]{\begin{eqnarray}\label{#1} }
\newcommand{\eea}{\end{eqnarray}}

\newcommand{\eqref}{\refb}

\usepackage{bm}
\usepackage[table]{xcolor}

\begin{document}

\baselineskip 24pt

\begin{center}

{\Large \bf Revisiting  localization  for BPS black hole entropy}

\end{center}

\vskip .6cm
\medskip

\vspace*{4.0ex}

\baselineskip=18pt

\centerline{\large \rm Ashoke Sen}

\vspace*{4.0ex}

\centerline{\large \it International Centre for Theoretical Sciences - TIFR 
}
\centerline{\large \it  Bengaluru - 560089, India}


\vspace*{1.0ex}
\centerline{\small E-mail:  ashoke.sen@icts.res.in}

\vspace*{5.0ex}

\centerline{\bf Abstract} \bigskip

Macroscopic computation of the degeneracy / index of a BPS black hole requires carrying
out string theory path integral over the near horizon geometry of the black hole. It has been
suggested that
the path integral can be localized to a finite dimensional subspace using the unbroken
supersymmetries of the near horizon geometry. We discuss several issues that need to be
resolved before one can declare victory: (1) delta-function normalizable eigenmodes of the
kinetic operator do not form a representation of the supersymmetry algebra 
since the Killing spinors grow exponentially as we approach the
asymptotic boundary of the near horizon geometry,  (2) the logarithmic contributions coming
from various one loop determinants are sensitive to the ultraviolet cut-off and therefore without
detailed knowledge of the ultraviolet cut-off one cannot determine the order one correction
to the entropy that could depend on the ratios of the charges, 
(3) the results of index theorem indicate that there are additional flat directions
of the localization action, at least at the quadratic order, than what has been considered so far.
We also discuss a limited application of localization techniques to  carry out the
path integral over the zero modes
of the metric and the gravitino, and get  a finite answer. This is consistent with the results of
arXiv:0905.2686 and shows that for computing BPS black hole entropy
it is not necessary to regulate the
path integral over these zero modes using the super-Schwartzian action.

\vfill \eject

\tableofcontents

\sectiono{Introduction} \label{sintro}

The index / degeneracy of supersymmetric black hole entropy can be computed
using supergravity -- or more generally string field theory --
path integral in the near horizon geometry of the 
black hole\cite{0809.3304}.
This provides a way to compare the microscopic and macroscopic results for the
index of such black holes\cite{9601029} beyond the leading term\cite{9711053,
0009234,0007195}. 
In particular, this
has been used
to compute and compare the logarithmic corrections to the black hole entropy on the
macroscopic and the microscopic side\cite{9709064,1005.3044,1106.0080,1108.3842,1109.3706,1311.6286,
1402.2441,1404.1379,1404.6363}.

On the microscopic side we have exact formula for black hole index in many cases
\cite{9607026,9903163,0505094,0605210,0607155,0609109}. However, the analysis of
the string field theory path integral that is needed for entropy computation
on the macroscopic side is hard to extend beyond the order to which it has been 
computed so far.  Taking inspiration from a similar suggestion for the world-sheet
integrals\cite{0608021} and gauge theories\cite{0712.2824,0904.4486}, it was suggested
in\cite{0905.2686,1012.0265,1111.1161,1208.6221,1404.0033,1504.01400,1504.01700,
1806.03690,1806.04479,1904.08876,
2209.13602,2211.06873} that we can use
the unbroken supersymmetry of the near horizon geometry of the black hole to localize
the path integral to a finite dimensional integral and carry out the
computation.\footnote{Localization has also been used to analyze quantum field theories
on $AdS_2\times S^1$\cite{1609.07443,1802.00427,1906.02722}.}
This has yielded results in agreement with the microscopic results under suitable
assumption on the integration measure over the supergravity fields. The latest
status of this approach can be found in \cite{2211.06873}.

There are, however, several issues that need to be understood better before one can claim
complete agreement between the microscopic and the macroscopic results.
\begin{enumerate}
\item 
There is  a possible obstruction to using supersymmetry
to localize the path integral over fields in the near horizon geometry\cite{1802.00427,1906.02722}. 
The near horizon geometry of the BPS black hole contains an euclidean $AdS_2$ 
factor that
describes a non-compact space. 
Typically while
integrating over the fields in a non-compact space, we need to impose appropriate boundary
condition on the fields, and the most natural boundary condition is to demand that the
fluctuations around the background must be square integrable. Square integrable fluctuations
of various tensor and spinor fields in $AdS_2$ may be expressed as a linear combination of
(delta function) normalizable eigenfunctions of the kinetic operator. 
On the other hand the Killing spinors that
describe the unbroken supersymmetry transformations in the near horizon geometry
are non-normalizable modes that grow exponentially as we approach the boundary of
$AdS_2$. When we transform a mode associated with one of the delta-function normalizable
basis functions by such a Killing spinor, we typically get a mode that does not fall off sufficiently
fast 
near the boundary of $AdS_2$ to render them normalizable. As a result, supersymmetry
generated by the Killing spinors do not seem to have a well defined action on the space
of (delta function) normalizable eigenmodes of the kinetic operator
and cannot be used to localize the path integral over
these modes. It may be possible to use a different basis\cite{1802.00427} or
non-normalizable boundary condition on the fields\cite{1806.04479}
to resolve this issue, 
but the relationship between such path integrals and path integral over
normalizable modes that is used to carry out the logarithmic correction to the black hole entropy
remains to be understood. Some discussion on boundary conditions 
may be found in \cite{1802.00427,2302.10370} and we
discuss some other possibilities in \S\ref{sdiscuss}. \label{p1}
\item Even if the problem mentioned above can be circumvented, and the path integral can be
localized on a finite dimensional manifold, we need to compute the one loop determinants of
the fluctuating fields representing directions transverse to the localization manifold. 
These determinants are ultraviolet  (UV) divergent and need to be regulated. The result depends
sensitively on the UV cut-off, in that it contains powers of the size of the near horizon
geometry measured in units of the cut-off, 
and an order one change in the cut-off leads to an order one change
in the entropy that could depend on the ratios of the charges.
In string theory one expects this cut-off to be provided by the string scale, but
without knowing the precise form of the cut-off one cannot determine the order one correction
to the entropy. In the existing calculation the cut-off is implicitly assumed to be the four
dimensional Planck scale.
\label{p2}
\item \cite{1806.04479,2209.13602} used an index theorem to compute the one loop 
determinant of fluctuations around
the localization manifold. The result of the index theorem indicates however that the determinant
is ill defined due to the presence of some zero eigenvalues. Therefore one needs to be more
careful in carrying out the path integral over the fields along these eigenmodes. \label{p3}
\end{enumerate}

We show that there is a small subset of normalizable modes that admit the action of
the unbroken supersymmetry without having to tamper with the boundary conditions.
These are the 
normalizable zero modes
of the metric and gravitino, generated by diffeomorphism / supersymmetry with
non-normalizable parameters. 
These transformations, together with the unbroken supersymmetries and
isometries of $AdS_2\times S^2$, generate the $\NN=4$ super-Virasoro algebra, and
hence the action of the unbroken supersymmetry on the zero modes can be found
from the $\NN=4$ super-Virasoro algebra. Therefore, despite the exponential growth of the
Killing spinors near the boundary of $AdS_2$, supersymmetry does have
well defined action within the
subspace of normalizable zero modes. We trace this to appropriate cancellation
between the exponentially growing terms. Therefore in this sector the unbroken
supersymmetry can be used to localize the path integral over the zero modes. Although 
we still have the zero eigenvalues mentioned in the point \ref{p3} above,
in this sector the corresponding eigenmodes span a bosonic
manifold of finite volume and hence integration over these modes yields a
finite result, confirming the result of \cite{0905.2686}. 
However the problem with the UV cut-off mentioned in point \ref{p2} above is
still present and hence the order one contribution remains ambiguous.
Recently this path integral was regularized by switching on a small
temperature that breaks the isometries of $AdS_2\times S^2$ geometry and produces
the super-Schwartzian action\cite{2209.13602,2209.13608}. 
While this remains a valid procedure and may be necessary
for the analysis of non-supersymmetric black hole entropy, this does not seem to be
needed for BPS black holes since localization already renders the result finite.

The rest of the paper is organized as follows.  In \S\ref{skilling} we review the 
construction of the Killing spinors that describe the unbroken supersymmetries of the
near horizon geometry of BPS black holes in $\NN=2$ supergravity in four dimensions
and verify that they grow exponentially as we approach the asymptotic region of $AdS_2$.
In \S\ref{sconflict} we review the (delta function) normalized basis functions for
various integer and half integer spin fields and show that a supersymmetry transformation by a 
Killing spinor changes (delta function) normalizable modes to non-normalizable modes
due to the exponential growth of the Killing spinor. This indicates that within the space of
normalizable fluctuations there is no well defined action of supersymmetry and hence
supersymmetry cannot be used to localize the path integral over the fluctuations of the
fields.  In \S\ref{snilp} we discuss the issue mentioned in point \ref{p3} above,
i.e.\  even if the
normalizability issue is circumvented,  index theorem indicates that there are
additional bosonic and fermionic modes, than what has been considered so far, over which we
need to carry out the integral explicitly. In \S\ref{suv} we discuss the issues related to
the UV cut-off.
In \S\ref{szero} we show that there is a small subsector
containing the normalizable zero modes of the metric and gravitino deformations where
these problems are absent. Within
this sector supersymmetry does have a well defined action and therefore the path integral over
these zero modes can
be localized to a finite dimensional integral and yields a finite result, as 
anticipated in \cite{0905.2686}.  In \S\ref{slog} we use localization to compute the
logarithmic correction to the black hole entropy due to the zero modes and show that
the result agrees with the result of direct computation in \cite{1106.0080,1108.3842}.
In \S\ref{sdiscuss} we discuss various possible ways one could try to remedy the
problem described in \S\ref{sconflict}, as well as additional issues that we need to address
in order to get a completely well-defined result for the black hole entropy from the supergravity
path integral.
Related discussions can be found in
\cite{1802.00427,2302.10370}, but to the best of our knowledge, a full understanding is still
awaited.

\sectiono{Killing spinors} \label{skilling}

The issues that we shall discuss arise for any BPS black hole whose near
horizon geometry preserves eight supersymmetries, but we shall illustrate this
for pure $\NN=2$ supergravity theory in four dimensions. This theory
contains a metric $g_{\mu\nu}$, a massless vector $A_\mu$ and a pair of spin
$3/2$ fields $\psi^\pm_\mu$, where we have suppressed the spinor indices of
$\psi_\mu^\pm$. The part of the Euclidean 
action involving bosonic fields $g_{\mu\nu}$ and
$A_\mu$ is given by,
\be
S=\int d^4 x \, \sqrt{\det g}\, \left[ R -{1\over 4} F_{\mu\nu} F^{\mu\nu} \right]\, , \qquad F_{\mu\nu}
=\p_\mu A_\nu - \p_\nu A_\mu\, .
\ee
This theory has an $AdS_2\times S^2$ solution describing the near horizon limit of a
BPS black hole of the form:\footnote{In this description the background electromagnetic
field is imaginary.
However this can be avoided by considering a magnetically charged black hole instead of
an electrically charged black hole.}
\be \label{emetsa}
ds^2 = a^2 (d\eta^2 +\sinh^2\eta \, d\theta^2) + a^2 (d\psi^2 +
\sin^2\psi \, d\phi^2), \qquad F_{\mu\nu} dx^\mu\wedge dx^\nu =-4\, i\, a^{-1}\, \sinh\eta\,
d\theta\wedge d\eta
\, .
\ee
We shall choose the vierbeins $e^a\equiv e^a_\mu dx^\mu$ and  the inverse vierbeins
$E_a\equiv E^\mu_a {\p\over \p x^\mu}$ to be 
\be\label{evier}
e^0 =  a \, \sinh\eta\, d\theta, \quad e^1 =  a\, d\eta, \quad 
e^2 =   a\, \sin\psi\, d\phi,
\quad e^3 =   a\, d\psi\, ,
\ee
\be \label{ederiv}
E_0 ={1\over   a \, \sinh\eta}{\p\over \p\theta}, \quad 
E_1 ={1\over   a}\, {\p\over \p\eta}, \quad E_2 = {1\over   a\, \sin\psi}\, {\p\over \p\phi},
\quad E_3 = {1\over   a}\, {\p\over \p\psi}\, .
\ee
The supersymmetry transformation is generated by a pair of parameters $\eps^\pm$,
with the supersymmetry transformation law of $\psi^\pm_\mu$ being
given by\cite{Ferrara:1976fu}, 
\be\label{esusygrav}
\delta \psi^\pm_\mu = D_\mu \eps^\pm \pm {i\over 8} \, 
F_{\rho\sigma} \gamma^{\rho\sigma} \, \gamma_\mu\eps^\pm\, ,
\ee
where we have ignored terms quadratic in the gravitino fields on the right hand side.
Therefore the Killing spinor equations, obtained by setting the right hand side
of \refb{esusygrav} to 0, are,
\be \label{ekilling}
D_\mu \eps^\pm \pm {i\over 8} \, 
F_{\rho\sigma} \gamma^{\rho\sigma} \, \gamma_\mu\eps^\pm=0 \, .
\ee

We shall use the $\gamma$-matrix convention\cite{1106.0080}:
\be \label{edefgamma}
\gamma^0 =- \sigma_3
\otimes \tau_2, \quad \gamma^1 =\sigma_3
\otimes \tau_1, \quad \gamma^2 =- \sigma_2
\otimes I_2, \quad \gamma^3 =\sigma_1
\otimes I_2, 
\ee
where $\sigma_i$ and $\tau_i$ are both Pauli matrices, but we have used different
symbols to make clear which space they act on.  We also define
\be \label{edefC}
\bar\eps \equiv
\eps^T C, \qquad C\equiv i\, \sigma_2\otimes \tau_1=\gamma^1 \gamma^3\, .
\ee
With this definition $\bar\eps_1 \eps_2$ can be shown to be Lorentz invariant for arbitrary
spinors $\eps_1,\eps_2$.
We use the contraction with $e_\mu^a$ or $E^\mu_a$ to convert  the space-time indices to tangent space indices and vice versa for
the $\gamma$
matrices, derivatives and various fields.
Using the action of covariant derivatives on spinors,
\be 
D_c = E_c^\mu D_\mu, \qquad 
D_\mu = \p_\mu +{1\over 4} \omega_\mu^{ab} \gamma^{ab}, \qquad
\gamma^{ab}\equiv {1\over 2} [\gamma^a,\gamma^b],
\ee
and the results,
\be
\omega_\phi^{23} =\cos\psi, \qquad \omega_\theta^{01} = \cosh\eta,
\ee
we get
\ben
&& D_0 = a^{-1} \, \left({1\over \sinh\eta}\p_\theta + {i\over 2}\, \tau_3\, \coth\eta\right), \quad D_1 = 
a^{-1}\p_\eta, \nonumber \\
&& D_2 = a^{-1} \,  \left({1\over \sin\psi}\p_\phi + {i\over 2}
\, \sigma_3\cot\psi\right),
\quad D_3 = a^{-1} \, \p_\psi\, .
\een
The Killing spinor equations \refb{ekilling} now take the form
\ben
&&a^{-1} \, \left({1\over \sinh\eta}\p_\theta + {i\over 2}\, \tau_3\, \coth\eta\right)\, 
\eps^\pm \mp {1\over 2} 
\, a^{-1}\, \sigma_3\otimes \tau_1 \, \eps^\pm =0, \nonumber \\
&& a^{-1}\p_\eta \,  \eps^\pm \mp {1\over 2} a^{-1} \, \sigma_3\otimes \tau_2 \, \eps^\pm =0\, ,
\nonumber \\
&&a^{-1} \, \left({1\over \sin\psi}\p_\phi + {i\over 2}\, \sigma_3\, \cot\psi\right)\, 
\eps^\pm \mp {i\over 2} \, a^{-1}\, \sigma_2\otimes \tau_3 \, \eps^\pm =0, \nonumber \\
&& a^{-1} \, \p_\psi \,  \eps^\pm \pm {i\over 2} a^{-1} \, \sigma_1\otimes \tau_3 \, \eps^\pm =0\, .
\een
There are eight solutions to these equations, given by,
\ben\label{ekillingspinor}
&& \eps^{\pm +}_- = e^{i(\phi-\theta)/2}\pmatrix{\sin{\psi\over 2} \cosh{\eta\over 2}\cr \pm i\, \cos{\psi\over 2}\cosh{\eta\over 2}\cr
\pm i\,
\sin{\psi\over 2}\sinh{\eta\over 2}\cr \cos{\psi\over 2} \sinh{\eta\over 2}}\, ,  \qquad
\eps^{\pm +}_+ = e^{i(\theta+\phi)/2}
\pmatrix{\sin{\psi\over 2} \sinh{\eta\over 2}\cr \pm i\, \cos{\psi\over 2}
\sinh{\eta\over 2}\cr
\pm i\,
\sin{\psi\over 2}\cosh{\eta\over 2}\cr 
\cos{\psi\over 2} \cosh{\eta\over 2}}\, , \nonumber \\ \\
&& \eps^{\pm -}_- = e^{-i(\phi+\theta)/2}\pmatrix{\cos{\psi\over 2} \cosh{\eta\over 2}\cr \mp i\, \sin{\psi\over 2}\cosh{\eta\over 2}\cr\pm i\,
\cos{\psi\over 2}\sinh{\eta\over 2}\cr 
-\sin{\psi\over 2} \sinh{\eta\over 2}}\, , \qquad
\eps^{\pm -}_+ = e^{i(\theta-\phi)/2}\pmatrix{\cos{\psi\over 2} \sinh{\eta\over 2}\cr \mp i\, \sin{\psi\over 2}\sinh{\eta\over 2}\cr\pm i\,
\cos{\psi\over 2}\cosh{\eta\over 2}\cr 
-\sin{\psi\over 2} \cosh{\eta\over 2}}\, . \nonumber
\een

Although the solutions appear to be non-periodic under $\theta\to \theta+2\pi$ or
$\phi\to\phi+2\pi$, and hence singular at the origin of $AdS_2$ and at the poles of $S^2$,
this can be traced to the fact that the vierbeins given in \refb{evier}, \refb{ederiv}
are themselves singular at these points. As discussed in \cite{0608021,0905.2686},
by appropriate choice
of non-singular vierbeins, we can make the Killing spinors manifestly single valued and
non-singular.

\sectiono{Supersymmetry and normalizability} \label{sconflict}

One notable property of the solutions \refb{ekillingspinor} to the Killing spinor equations
is that for large $\eta$ they grow as $e^{\eta/2}$.
Due to this, supersymmetry transformation by a Killing spinor converts a normalizable
fluctuation to a non-normalizable fluctuation and therefore these supersymmetries do not
act within the space of (delta function)
normalizable modes of fluctuation satisfying standard boundary
conditions.  In the rest of the section we shall elaborate on this
observation. A different version of this problem was discussed in \cite{1802.00427} where it was 
shown that supersymmetry transformation by a Killing spinor can some time convert a
normalizable solution to the equations of motion to a non-normalizable solution.

For convenience, we shall work with fields carrying tangent space indices only, so that
while computing the norm,
the contraction with the metric does not produce any extra
dependence on $\eta$.
If a particular mode of a 
field carrying tangent space indices grow as $e^{\alpha\eta}$ for
large $\eta$, we define the degree of growth of the mode to be 
$\alpha$. Since the volume element of $AdS_2$, given by 
$\sinh\eta \, d\eta \, d\theta$, grows as $e^\eta$, nornalizable modes
should have $\alpha<-1/2$. On the other hand delta function normalizable modes, carrying 
factors of 
\be\label{easymp}
e^{-\eta/2} e^{\pm i\lambda\eta}\, ,
\ee 
for large $\eta$, 
have degree of growth $-1/2$, so that the
inner product between two such modes takes the form
\be
\int d\eta \, \sinh\eta\, e^{-\eta} e^{i\eta (\lambda-\lambda')} \propto \delta(\lambda-\lambda')\,.
\ee
For future reference we note that the derivative operators $E_1,E_2$ and $E_3$
defined in \refb{ederiv} preserves the degree of growth of a function while $E_0$
decreases the degree of growth by 1.

We shall first verify \refb{easymp} for various known basis states of 
different types of fields in $AdS_2\times S^2$\cite{campo,camhig1,campo2,camhig2,9505009}, given by the eigenstates of the
Klein-Gordon or Dirac operators and their derivatives.
For example the eigenfunctions of the scalar Laplacian in $AdS_2$,
with eigenvalue $-\left({1\over 4}+\lambda^2\right)/a^2$, are given 
by
\ben \label{e5pa}
f_{\lambda,k}(\eta,\theta)
&=& e^{ik\theta} \sinh^{|k|}\eta\, F\left(i\lambda +{1\over 2}+|k|, -i\lambda 
+{1\over 2}+|k|; |k|+{1}; -\sinh^2{\eta\over 2}\right), \nn
&& \qquad \qquad \qquad \qquad
k\in \ZZZ, \qquad 0<\lambda<\infty\, .
\een
Here $F\equiv ~_2F_1$ denotes hypergeometric function.  Product of 
$f_{\lambda,k}$ and the spherical harmonics $Y_{\ell,m}$ on $S^2$ provide a complete set
of basis states for scalar fields in $AdS_2\times S^2$.
Using the asymptotic growth formula,
\be\label{ehypergrowth}
F(a,b,c;z) \sim z^{-a} \frac{(-1)^{-a} \Gamma (b-a) \Gamma (c)}{\Gamma (b) \Gamma
   (c-a)} +z^{-b}\frac{(-1)^{-b}
   \Gamma (a-b) \Gamma (c)}{\Gamma (a) \Gamma
   (c-b)} \qquad \hbox{for large $|z|$},
\ee
we see that for large $\eta$, \refb{e5pa} grows as
\be
e^{-\eta/2} e^{\pm i\lambda\eta}\, ,
\ee
following the requirement of delta function normalizability. 

Let us now turn to spinor fields.
If we define the Dirac operator on $AdS_2$ and $S^2$ as,
\be
\not\hskip -4pt D_{AdS_2} = -\tau_2 D_0 +\tau_1 D_1, \qquad
\not\hskip -4pt D_{S^2}=-\sigma_2 D_0 +\sigma_1 D_1,
\ee
then the eigenstates of $\not\hskip -4pt D_{AdS_2}$ take the form\cite{campo2}
\ben \label{ed2aa}
\chi_{ k}^{\pm}(\lambda) = 
e^{ i\left( k+{1\over 2}\right)\theta}  \pmatrix{ i \, {\lambda}\, 
\cosh^{ k}{\eta\over 2}\sinh^{ k+1} {\eta\over 2}
F\left( k+1+i\lambda,  k+1-i\lambda;  k+2;-\sinh^2{\eta\over 2}\right)
\cr
\pm ( k+1)\, \cosh^{ k+1}{\eta\over 2}\sinh^ k {\eta\over 2}
F\left( k+1+i\lambda,  k+1-i\lambda;  k+1;-\sinh^2{\eta\over 2}\right)}, \nn \cr \cr
\eta_{ k}^{\pm}(\lambda) = 
e^{ -i\left( k+{1\over 2}\right)\theta}
\pmatrix{ ( k+1)\, \cosh^{ k+1}{\eta\over 2}\sinh^ k {\eta\over 2}
F\left( k+1+i\lambda,  k+1-i\lambda;  k+1;-\sinh^2{\eta\over 2}\right)
\cr
\pm i \, {\lambda}\, 
\cosh^{ k}{\eta\over 2}\sinh^{ k+1} {\eta\over 2}
F\left( k+1+i\lambda,  k+1-i\lambda;  k+2;-\sinh^2{\eta\over 2}\right)
}, 
\nn \cr 
  k\in \ZZZ, \quad 0\le  k<\infty, \quad 0<\lambda<\infty\, , \hskip 1in
\een
and satisfy
\be \label{eadseva}
\not \hskip -4pt D_{AdS_2} \chi_{ k}^{\pm}(\lambda)=\pm i\, a^{-1}\,
\lambda\,  \chi_{ k}^{\pm}(\lambda)
\, , \qquad
\not \hskip -4pt D_{AdS_2} \eta_{ k}^{\pm}(\lambda)
=\pm i\, a^{-1}\,
\lambda\,  \eta_{ k}^{\pm}(\lambda)
\, .
\ee
Similarly, eigenstates of  $\not\hskip -4pt D_{S^2}$ take the form:
\ben \label{ed2bb}
\chi_{l,m}^{\pm} &=& {\sqrt{(l-m)!(l+m+1)!}\over l!}\,
e^{ i\left(m+{1\over 2}\right)\phi} 
\pmatrix{ i\, \sin^{m+1}{\psi\over 2}\cos^m {\psi\over 2}
P^{\left(m+1, m\right)}_{l-m}(\cos\psi)
\cr
\pm \sin^{m}{\psi\over 2}\cos^{m+1} {\psi\over 2}
P^{\left(m, m+1\right)}_{l-m}(\cos\psi)
}, \nn
\eta_{l,m}^{\pm} &=& 
{\sqrt{(l-m)!(l+m+1)!}\over l!}\,
e^{ -i\left(m+{1\over 2}\right)\phi} 
\pmatrix{ \sin^{m}{\psi\over 2}\cos^{m+1} {\psi\over 2}
P^{\left(m, m+1\right)}_{l-m}(\cos\psi)
\cr
\pm i \, \sin^{m+1}{\psi\over 2}\cos^m {\psi\over 2}
P^{\left(m+1, m\right)}_{l-m}(\cos\psi)}, 
\nn
&& 
\qquad l,m\in \ZZZ, \quad l\ge 0, \quad 0\le m\le l\, ,
\een
where
\be \label{ed4aa}
P_n^{(\alpha,\beta)}(x) = { (-1)^n\over 2^n \, n!} (1-x)^{-\alpha}
(1+x)^{-\beta} {d^n\over dx^n} \left[ (1-x)^{\alpha+n}
(1+x)^{\beta+n}\right]\, .
\ee
These eigenstates satisfy
\be \label{ed3aa}
\not \hskip -4pt D_{S^2} \chi_{l,m}^\pm =\pm i\, a^{-1}\,
\left(l +1\right) 
\chi_{l,m}^\pm\, , \qquad
\not \hskip -4pt D_{S^2} \eta_{l,m}^\pm =\pm i\, a^{-1}\,
\left(l +1\right) 
\eta_{l,m}^\pm\, .
\ee
Product of eigenfunctions of 
$\not\hskip -4pt D_{AdS_2}$ and $\not\hskip -4pt D_{S^2}$ provide a complete set
of basis vectors for the expansion of spinor fields in $AdS_2\times S^2$.
One can again verify, using \refb{ehypergrowth}, that \refb{ed2aa} grow as
$e^{-\eta/2} e^{\pm i\lambda\eta}$ for large $\eta$, thereby satisfying the condition
for delta function normalizability. 

The basis states for higher spin fields can be constructed from the basis states
of the scalars and fermions given above as follows. First consider vector fields.
Basis states for vector fields in $AdS_2$ are given by $\{ d f_{\lambda,k}\}$ and
$\{* d f_{\lambda,k}\}$ where $*$ is the Hodge dual
in $AdS_2$, and an additional set of discrete
modes that will not be relevant for the discussion of this section. It can be easily checked
that once we use tangent space indices to describe these vectors, they grow as
$e^{-\eta/2} e^{\pm i\lambda\eta}$ for large $\eta$ and therefore remain delta function
normalizable. Similarly the
basis states of vector fields on $S^2$ are given by $dY_{\ell,m}$ and $\star dY_{\ell,m}$
except for $\ell=0$ for which $dY_{\ell,m}=0$. Here $\star$ denotes the Hodge dual in
$S^2$. The basis states of vector fields on $AdS_2\times S^2$ are then constructed from
the product of the basis states for vectors in $AdS_2$ and the basis states for 
scalars in $S^2$ or
vice versa.

The basis states for metric deformations in $AdS_2$ can be taken to be proportional
to $g_{mn} f_{\lambda,k}$ and $D_m \xi_n + D_n \xi_m - 2 g_{mn} D^p\xi_p$ where $g_{mn}$
is the $AdS_2$ background metric, $f_{\lambda,k}$ is one of the basis states for scalars
given in \refb{e5pa} and $\xi$ is one of the basis states for vector fields in $AdS_2$ described
above. Besides these, there are also a set of discrete modes that will not be relevant for
discussion in this section. The basis states for metric deformation in $S^2$ can be
constructed similarly. Basis states for metric deformation in $AdS_2\times S^2$ are 
obtained by taking the product of the  basis states
for vector fields on $S^2$ and the basis states for vector fields on $AdS_2$, product of the basis
states for metric deformation in $AdS_2$ times the basis states of scalars in $S^2$ and the
product of the basis states for metric deformation in $S^2$ and the basis states for scalars in
$AdS_2$.

Finally, for every basis state for spinors, we can construct four basis
states for the Rarita-Schwinger field by applying either the $\gamma^\mu$'s or
the $D_\mu$'s
on the spinor basis. For example, a possible choice of basis, 
used {\it e.g.} in \cite{1106.0080,1108.3842}, is
\be\label{egravitinobasis}
(\gamma_0,\gamma_1,0,0)\chi, \qquad (D_0,D_1,0,0)\chi, \qquad (0,0,\gamma_2,
\gamma_3)\chi, \qquad
(0,0,D_2,D_3)\chi\, .
\ee
Besides these, there are also discrete modes that will not be important for the discussion
of this section. 

The path integral of the gauge fixed theory also includes integration over the ghost fields and
we need to introduce complete set of basis states for them as well. These can be constructed
in a manner similar to those described above.

The actual eigenmodes of the kinetic operator of the gauge fixed theory can be constructed
as finite linear combinations of the modes described above\cite{1005.3044,1106.0080,
1108.3842}.
It follows from the asymptotic growth of the basis states of scalars and spinors in $AdS_2$
that the delta function normalizable basis states of all the fields, carrying tangent
space indices, grow as $e^{-\eta/2}e^{\pm 
i\lambda\eta}$ for large $\eta$.
Once the basis states are given, we can expand  various fields as linear
combination of the basis states and carry out integration over the coefficients of
expansion to compute the supergravity partition function in the near horizon geometry.
This in turn is expected to produce the black hole degeneracy / index in the microcanonical
ensemble\cite{0809.3304}. 
The logarithmic corrections to BPS black hole entropy were computed by
following precisely this procedure\cite{1005.3044,1106.0080,1108.3842}.\footnote{For
carrying out localization of the path integral using supersymmetry, one needs to have
a closure of the supersymmetry algebra off-shell. This often requires introduction
of new auxiliary fields. However the original fields like metric and gravitino in the gravity
multiplet, vectors, scalar and spinors in the vector multiplet and spinors and scalars in
the hypermultiplet are still present in the off-shell completion and the basis states
introduced here can continue to be used for mode expansions of these fields.}

Now consider a particular mode of fluctuation $\hat A_\mu$
of the gauge field $A_\mu$ on $AdS_2\times S^2$
of the form
\be
\hat A_\mu dx^\mu = u\, *df_{\lambda,k} \, ,
\ee
where $u$ is an arbitrary coefficient over which we need to integrate for computing the
partition function of $AdS_2\times S^2$. The fluctuation in the field strength 
$\hat F_{\mu\nu}$
is given by,
\be
{1\over 2} \hat F_{\mu\nu} dx^\mu \wedge dx^\nu = u\, d*df_{\lambda,k} 
= - u\, * * d*df_{\lambda,k}  =- u\, * \square f_{\lambda,k} 
=  u\, \left({1\over 4}+\lambda^2 \right) \, a^{-2} * f_{\lambda,k}\, ,
\ee
where we used the fact that in two dimensions $**=-1$ and, acting on scalars,
$* d * d = \square$. This gives, in tangent space,
\be
\hat F_{01} = u\, \left({1\over 4}+\lambda^2 \right) \, a^{-2}  f_{\lambda,k}\, .
\ee

Now consider the supersymmetry transformation of the gravitino given in \refb{esusygrav}
under any of the Killing spinors in the presence of the mode described above. Since the
background contribution to the right hand side of \refb{esusygrav} vanishes, the
only non-vanishing contribution comes from the fluctuations $\hat F_{\mu\nu}$ and we get
\be
\delta \psi^\pm_\mu =\pm {i\over 8} \, 
\hat F_{\rho\sigma} \gamma^{\rho\sigma} \, \gamma_\mu\eps^\pm
=\mp {1\over 4} \, u\, \left({1\over 4}+\lambda^2 \right) \, a^{-2}  f_{\lambda,k} 
\tau_3 \, \gamma_\mu \eps^\pm\, .
\ee
Converting this to the tangent space notation we get
\be\label{e317}
\delta \psi^\pm_b = \mp {1\over 4} \, u\, \left({1\over 4}+\lambda^2 \right) \, a^{-2}  f_{\lambda,k} 
\tau_3 \, \gamma_b \eps^\pm\, , \qquad 0\le b\le 3\, .
\ee
Using the growth rate $f_{\lambda,k}\sim e^{-\eta/2} e^{\pm i\lambda \eta}$ and
$\eps^\pm \sim e^{\eta/2}$ for large $\eta$, we see that $\delta \psi^\pm_b\sim 
e^{\pm i\lambda\eta}$ for large $\eta$, and therefore cannot be expressed as a
linear combination of delta function normalizable modes. Furthermore 
linear combination of the Killing spinors cannot be used to resolve this problem,
since they have different dependence on the $\theta,\phi$ coordinates.
Since the partition
function in $AdS_2\times S^2$ is computed by carrying out path integral over
(delta function) normalizable modes, it would seem that supersymmetry generated
by the Killing spinors  cannot be used to localize this path integral.\footnote{In supergravity,
the symmetry that is used for the localization of the path integral is a linear combination
of the usual supersymmetry and the BRST transformation\cite{1806.04479}.
But use of such a linear
combination does not help resolve the normalizability issue as long as each of the
fields is allowed to fluctuate independently of the others.}
We shall see another
example of this in \S\ref{szero}.

Even though we have illustrated the problem using a specific example, this problem
is pervasive, and generically, supersymmetry transformation by a Killing spinor of any
delta function normalizable mode produces  a mode that is not delta function 
normalizable and is therefore not a linear combination of the modes over which we
integrate. The same problem persists when we add vector and / or hypermultiplets
to the theory, and in each supermultiplet, the effect of transformation by a Killing spinor
converts a (delta function) normalizable mode to a non-normalizable mode.

Similar problem also appears for the discrete modes of the gauge fields, metric and
gravitino fields. These modes are given by the product of (derivatives of) $Y_{\ell m}$ and
a set of basis functions in $AdS_2$ that are obtained from the ones given above by setting
$\lambda$ to special imaginary values. They fall off faster than $e^{-\eta/2}$ for large $\eta$ and
are therefore normalizable. However their supersymmetry transformations often produce
modes that are not part of the spectrum of normalizable modes, especially if the part of the
function associated with $S^2$ has $Y_{\ell m}$ for generic $\ell>1$ whose derivative does
not vanish or satisfy special relations.

In \S\ref{szero} we shall see an exception to this rule for integration over
the zero modes of the metric and the gravitino.

One could ask if the problem discussed here
can be mitigated with the help of a local Lorentz transformation.
For example, \cite{1810.11442} found that the Killing spinor in 
$AdS_5$, that grow exponentially like the ones
in $AdS_2$, can be made to grow like a constant by changing the local Lorentz frame.
This of course requires use of local Lorentz transformations that grow exponentially at
infinity, but unlike gauge, general coordinate and supersymmetry transformations, local
Lorentz transformations may be declared as pure gauge even if they blow up at infinity.
One might wonder if something similar can be done here. It is easy to prove that there is no
choice of frame in which all the Killing spinors have zero degree of growth as $\eta\to\infty$.
For this we can check that the invariant combination $\bar \eps_1\eps_2=\eps_1^T C\eps_2$
has degree of growth 1 for some choices of Killing spinors, {\it e.g.}  
one can check using \refb{ekillingspinor} and \refb{edefC} that 
$\bar\eps^{++}_+\eps^{--}_-$ has degree of growth 1 and hence we cannot choose a frame
in which $\eps^{++}_+$ and $\eps^{--}_-$ can both be made to have degree of growth 0.
Since localization uses the combination of transformations generated by 
$\eps^{++}_+$ and $\eps^{--}_-$ (or some other similar pairs), it does not appear
that just by changing the Lorentz frame
we can resolve the normalizability issue.

In \S\ref{sdiscuss} we shall discuss, but will not fully resolve, the issues discussed in this
section.

\sectiono{$L_0=J_0$ sector} \label{snilp}

In this section we shall discuss another issue that arises in applying localization to
supergravity path integrals, assuming that the issues raised in the last section can be
resolved, possibly by choosing a different basis of functions. 
Let us briefly
recall the basic idea of localization in the context of supergravity black holes, as used
in \cite{0905.2686,1012.0265,1111.1161,1208.6221,
1404.0033,1504.01400,1504.01700,
1806.03690,1806.04479,1904.08876,
2209.13602,2211.06873}. We extend the space of supergravity
fields by introducing sufficient number
of auxiliary fields and ghosts so that acting on these fields, 
a particular linear combination $Q$ of
supersymmetry and BRST operators satisfies
\be
Q^2 = L_0-J_0\, ,
\ee
where $L_0$ and $J_0$ generate $\theta$ and $\phi$ translations respectively. 
We can add to the original action a term
\be\label{elocaction}
t\, Q\VV, \qquad \VV=\int \sum_a (Q\Psi_a(x))^* \Psi_a(x)\, ,
\ee
where the sum over $a$ runs over all the fermionic fields in the theory and $t$ is an arbitrary
constant. Since $Q$
is a symmetry of the original action, the addition of this term does not change the
partition function as long as $\VV$ is invariant under $L_0-J_0$. 
The bosonic
part of this action \refb{elocaction} takes the form
\be
t\, \int \sum_a (Q\Psi_a(x))^* (Q\Psi_a(x))\, .
\ee
Since this is positive definite, as $t\to \infty$, this localizes the path integral to the
subspace $Q\Psi_a=0$ for every $a$.

In the sector with non-zero $(L_0-J_0)$ eigenvalue $h$, 
the bosonic and fermionic modes are
paired as\footnote{As will be discussed in \S\ref{slog}, there will be additional powers
of the $AdS_2$ size parameter $a$ in these equations if we use correctly normalized
modes, but for the discussion in this section we can ignore this effect.}
\be\label{earep}
Q\, \phi_f= \phi_b, \qquad Q\, \phi_b =  h\, \phi_f\, ,
\ee
at linearized level. In the $t\to\infty$ limit, the path integral over these modes just produces
the one loop determinant of the kinetic operators computed from the action \refb{elocaction}.
However in the $(L_0-J_0)=0$ sector we need to be more careful.
In the basis described in \S\ref{sconflict} these correspond to the modes for which $k=m$,
and will be labelled by $\lambda$ and $\ell$. In a different basis with different boundary
conditions there will be different labels, but we still expect infinite number of such modes.
In this sector the modes may be divided into three classes: (a) modes $\phi$ 
that are
not annihilated by $Q$, (b) modes $\tilde\phi=Q\phi$ that are annihilated by $Q$ 
and (c) modes that are annihilated by $Q$ but are not of type (b). Of these
type (a) and (b) modes are paired. 
Type (c) modes are not paired.
We add a superscript $b$ or $f$ to the modes $\phi$, $\tilde\phi$
denote if the mode is bosonic or fermionic,
{\it e.g.} we collectively denote a bosonic mode of type (a) by $\phi_b$ and their fermionic
partners by $\tilde\phi_f=Q\phi_b$. 
Also we shall include in the list of $\phi_b$ and $\phi_f$ the modes of type (c),
and regard $\phi_b,\phi_f,\tilde\phi_b,\tilde\phi_f$ as column vectors,
keeping in mind that some of the entries in $\tilde\phi_b=Q\phi_f$ and $\tilde\phi_f=Q\phi_b$
may vanish
if their partners in $\phi_f$ and / or $\phi_b$ represent modes of type (c). 
The relevant part of the action \refb{elocaction} involving these variables takes the form:
\be\label{eloczero}
t\, Q\, \VV, \qquad \VV = \pmatrix{\tilde\phi_f^T & \phi_f^T} 
\pmatrix{D_{00} & D_{01}\cr D_{10} & D_{11}} \pmatrix{\phi_b \cr \tilde\phi_b}\, ,
\ee
where the superscript $T$ denotes transpose and
$D_{ij}$ are some matrices, of which $D_{00}$ and $D_{11}$ are square matrices
and $D_{10}$ and $D_{01}$ are, in general, rectangular matrices. Using the relation 
$Q^2=0$, and the results:
\be
\tilde \phi_f = Q\phi_b, \qquad \tilde\phi_b = Q\phi_f\, ,
\ee
we get\footnote{I wish to thank Imtak Jeon and Gustavo Turiaci for explaining this 
construction.}
\ben\label{eloccomp}
&& t\, Q\, \VV = t \pmatrix{0 & \tilde\phi_b^T} \pmatrix{D_{00} & D_{01}\cr D_{10} & D_{11}} \pmatrix{\phi_b \cr \tilde\phi_b}
- t \pmatrix{\tilde\phi_f^T & \phi_f^T} \pmatrix{D_{00} & D_{01}\cr D_{10} & D_{11}} \pmatrix{
\tilde\phi_f \cr 0} \nonumber \\
&=& t \, \tilde\phi_b^T D_{10}\phi_b + {1\over 2} t \tilde\phi_b^T (D_{11}+D^T_{11}) \tilde\phi_b
- t\, \phi_f^T  D_{10} \tilde\phi_f - {1\over 2} t \tilde\phi_f^T (D_{00}-D^T_{00}) \tilde\phi_f\, .
\een
Now, if the kernel of $D_{10}$ is $n$ dimensional, then $D_{10}$ has $n$ right eigenvectors
with zero eigenvalue. For every such eigenvector we can identify a particular combination
of the modes $\phi_b$ that does not enter the action \refb{eloccomp}. Similarly, if the
cokernel of $D_{10}$ is $m$ dimensional, then $D_{10}$ has $m$ left eigenvectors
with zero eigenvalue. For every such eigenvector we can identify a particular combination
of the modes $\phi_f$ that does not enter the action \refb{eloccomp}. Therefore, integration
over such modes is not localized and we must carry out these integrals exactly by taking
into account the dependence of the original action on these modes. If $\phi_b$ and / or $\phi_f$
contains modes of type (c), then the column vectors
$\tilde\phi_f$ and / or $\tilde\phi_b$ have zero entries, and we may have additional linear
combinations of the modes $\phi_f$ and / or $\phi_b$ that decouple from the action
\refb{eloccomp}, besides the ones associated with the cokernel and the kernel of $D_{10}$.

The combination $(n-m)$ can be computed using an index theorem and has been
computed in \cite{1806.04479,2209.13602} for various multiplets for different values
of $(L_0-J_0)$. Our
interest is in the $(L_0-J_0)=0$ sector. From the results for the index for the
graviton multiplet of the $\NN=2$ supergravity, given {\it e.g.} by the coefficient of the
$q^0$ term in eq.(4.58) of 
\cite{1806.04479} or (4.44) of \cite{2209.13602},\footnote{The results in \cite{2209.13602}
are given only for the north pole  contribution. There will be a similar contribution from the
south pole.} we find that $n-m=4$. Therefore
there are at least four bosonic modes that do not enter the action \refb{eloccomp} and
must be integrated separately. We shall see in \S\ref{szero}, \S\ref{slog} that two of these modes can
be identified as part of the metric zero modes corresponding to the generators 
$J^+_{-1}$ and $J^-_1$ of the asymptotic symmetry group and integral over these modes
need to be evaluated
separately, yielding a contribution given in \refb{elocz}. 
This leaves us with at least two other modes, that
have not been identified so far, over which we must perform the integral
explicitly. Similarly eq.(4.45) of 
\cite{2209.13602} shows that the gravitino multiplet of $\NN=2$ supergravity has
$(n-m)=-4$, and therefore there are at least four fermionic modes that do not
enter the action \refb{eloccomp}. Since the $\NN=8$ supergravity, written in the
language of $\NN=2$ supergravity, has six such gravitino multiplets, we get twenty four
modes of this type and must explicitly carry out integration over these modes.

\sectiono{Ultraviolet cut-off} \label{suv}

In the computation of the one loop determinant around the localization manifold one
encounters expressions involving the logarithm of the size of the horizon. Since the size
is a dimensionful parameter, there must be some other length scale with respect to which
the size is measured. This is the UV cut-off of the theory. This can also be related to
the choice of the path integral measure.\footnote{In string theory we have a natural UV
cut-off. A related fact is that the path integral measure over the string fields is fixed by the
Batalin-Vilkovisky master equation.} This way, the UV cut-off
of theory enters the expression for the entropy.
For example \cite{2209.13602} 
implicitly takes this
cut-off to be the four dimensional Planck scale. 
However, in a 
UV complete theory like string theory, 
the natural cut-off is the string scale.
The difference between the two choices is an order one effect on the black hole entropy
that could depend on the ratios of the charges carried by the black hole.
Therefore if in the analysis of \cite{2209.13602} we take the cut-off to be the string scale,
we shall get the wrong result for the entropy. More generally, different fields could have
different effective UV cut-off, which could even depend on the local scalars constructed
from the background fields. Since the background fields depend on the charges carried
by the black hole, this could introduce non-trivial dependence on the black hole
charges. Even if we insist on using the duality invariant Einstein metric
$g_{\mu\nu}$ for fixing the cut-off, a field redefinition, {\it e.g.} of the form $g_{\mu\nu}\to
g_{\mu\nu}+R_{\mu\nu}$, could effectively change the cut-off and yield the wrong answer.
These issues are related to the general problem that while the
logarithm of a large variable (size of $AdS_2$ in this case) is easy to extract from an
integral,  finite and power
suppressed contributions to the integral depend sensitively on the end points of
integration.

This is not a purely academic issue. In the closely related example of BPS black holes in
$\NN=4$ supersymmetric theories,  there are no logarithmic corrections to the entropy
but there are order one corrections that depend non-trivially on the ratios of
charges\cite{9607026,0605210,0607155,0609109}.
These are reproduced by Riemann$^2$ corrections to the effective
action\cite{9711053,0009234,0007195,0510147,0607155,0609109}, 
but the 
computation of the effective action requires stringy effects and modular invariance in an
essential way\cite{9610237,9708062}. 
Even though there are no Riemann$^2$ correction of similar type to
the effective action of $\NN=8$ supersymmetric string theories, one would expect that
in any duality frame the natural UV cut-off will be provided by the string scale (or eleven
dimensional Planck scale if we are using the M-theory description) in that duality
frame, and at the end, after combining all terms, we should get a duality invariant result.

From this discussion it would seem that a fully satisfactory 
implementation of the
localization program can be achieved only after taking into account the UV complete
description of the theory.

\sectiono{Zero modes and their supersymmetry transformation} \label{szero}

The graviton and gravitino fields in $AdS_2\times S^2$ have a set of zero modes that
decouple from the rest of the modes. These can be regarded as the result of 
diffeomorphism and supersymmetry transformation with non-normalizable parameters
although the modes themselves are normalizable. 
While computing logarithmic correction to the black hole entropy we expand the action to
quadratic order in the fields, and the action does not depend on the zero modes. Therefore
they cannot be treated via gaussian integration.
It was argued in \cite{0905.2686} that localization
can be used to convert the infinite dimensional integral over these modes into a finite
dimensional integral leading to a finite answer. In this section we shall briefly recall the
argument and illustrate why localization can be used for evaluating the integration over the
zero modes despite the Killing spinors having positive degree of growth.

Before getting into the details of the analysis, let us briefly recall the importance of the zero
modes. Since the size $a$ of $AdS_2$ provides an effective infrared cut-off, while
computing the partition function of string theory in the near horizon geometry, we can encode
the effect of quantum fluctuations due to all the modes other than the
zero modes into an effective Lagrangian density $\LL_{\rm eff}$ and represent
the partition function as, 
\be
Z = \exp\left[-a^2\int_0^{\eta_0} d\eta \int_0^{2\pi} d\theta \, \sinh\eta\, d\eta \, d\theta \, \LL_{\rm eff}
+\hbox{boundary terms}\right]\, ,
\ee
where $\eta_0$ is a cut-off on the $\eta$ integral.    Furthermore the $SL(2,R)$ isometry of
$AdS_2$ ensures that $\LL_{\rm eff}$ evaluated in the background geometry is constant on 
$AdS_2$. Therefore we can express $Z$ as
\be\label{ezexp}
Z = \exp\left[-2\pi a^2 \, \LL_{\rm eff}\, (\cosh\eta_0-1) + \hbox{boundary terms}\right]\, .
\ee
Now the length $\beta$ of the $AdS_2$ boundary, which we can interpret as the inverse
temperature of the boundary observer, is given by
\be
\beta = 2\, \pi \, a\, \sinh\eta_0 = 2\pi a \, \cosh\eta_0 + \OO(e^{-\eta_0})\, .
\ee
Ignoring the order $e^{-\eta_0}$ terms, and noting that the contribution due to the boundary
terms is proportional to the boundary length $\beta$, we can express \refb{ezexp} as
\be\label{ezbound}
Z = A \, e^{-\beta E}\, ,
\ee
where
\be 
A= \exp\left[2\pi a^2 \, \LL_{\rm eff}\right], \qquad E = a \, \LL_{\rm eff} +\hbox{boundary
contribution}\, .
\ee
\refb{ezbound} shows that 
the boundary observer sees a system with degenerate spectrum, with degeneracy $A$
and energy $E$. $A$ can now be interpreted as the ground state 
degeneracy of the black hole\cite{0809.3304,1008.3801}.\footnote{There can also be
other saddle points with the same asymptotic geometry as the original near horizon
geometry, {\it e.g.} orbifolds of the original geometry\cite{0810.3472,0905.2686}, 
that contribute to the partition
function. The contribution from these saddles take the form $A' e^{-\beta E}$ with the same
constant $E$ that multiplies $\beta$ but different constants $A'$. Therefore in the final
formula for the degeneracy, we need to add the contributions $A$, $A'$ etc.}

The only missing part in this analysis is the contribution from the zero modes, which may
not necessarily be captured into an effective Lagrangian density $\LL_{\rm eff}$. They will
in general give additional contribution to $A$. If this contribution vanishes or diverges, then
the statistical interpretation of the system given above breaks down. Therefore it is
important to show that the zero modes give a finite contribution to $A$.

The zero modes of the graviton
are of the form\cite{camhig1}:
\ben \label{ehmna}
h_{\mu\nu} dx^\mu dx^\nu &=&
2\, k\, (k^2-1) \, {(\sinh\eta)^{|k|-2} \over (1 +\cosh\eta)^{|k|}}
\, e^{ik\theta}\, (d\eta^2 + 2 \, i\, \sinh\eta\, d\eta \, d\theta
- \sinh^2\eta \, d\theta^2) \nn
&& \quad k\in \ZZZ, \quad |k|\ge 2\, ,
\een
and
\ben \label{e24pa}
&& h_{\mu\nu} dx^\mu dx^\nu =  (\star d Y_{1m}(\psi,\phi))_{(\mu} \p_{\nu)} \Phi_k(\eta,\theta), 
\qquad \Phi_k = 
\left[ {\sinh\eta \over 1+\cosh\eta}\right]^{|k|} e^{ik\theta}, \nonumber\\
&& \hskip 1in m=0,\pm 1, \quad k = \pm 1, \pm 2, \pm 3, \cdots\, .
\een
Note that even though $\Phi_k$ is a non-normalizable mode of a scalar, 
$E_a^\mu\p_\mu\Phi_k$ has degree of growth $-1$ and hence
is a normalizable mode of a vector field. $\star dY_{1m}$
represent conformal Killing vectors on $S^2$. Both sets of zero modes may be
expressed as
\be\label{ediffzero}
h_{\mu\nu} = D_\mu\xi_\nu + D_\nu \xi_\mu\, ,
\ee
where
\ben\label{ediffpar1}
a\, (E_0^\mu  \xi_\mu,\cdots, E_3^\mu\xi_\mu)&=& 
e^{ik\theta} \, {(\sinh\eta)^{|k|-1} \over (1 +\cosh\eta)^{|k|}}  \nonumber \\
&& \hskip .5in  \left({i} \, 
\left(k^2 +\sinh^2\eta +|k|\cosh\eta\right) , k(\cosh\eta+|k|)
, 0,0\right)  \, ,
\een
for \refb{ehmna} and 
\be\label{ediffpar2}
a\, (E_0^\mu  \xi_\mu,\cdots, E_3^\mu\xi_\mu) = {1\over 2} \left(0,0, \p_\psi Y_{1m}(\psi,\phi),
-{1\over \sin\psi}\, \p_\phi Y_{1m}(\psi,\phi)
\right) \, \Phi_k(\eta,\theta)\, ,
\ee
for \refb{e24pa}.
The modes $E_c^\mu \xi_\mu$ given in \refb{ediffpar1}
have degree of growth 0 for $c=1$ and $1$ for
$c=0$. On the other hand the  modes $E_c^\mu \xi_\mu$ given in \refb{ediffpar2}
have degree of growth 0 for $c=2,3$.
Therefore $\xi_\mu$ are not normalizable and hence 
$D_\mu\xi_\nu + D_\nu \xi_\mu$ are not pure gauge deformations.
However $E_a^\mu E_b^\nu h_{\mu\nu}$ have degree of growth $\le -1$ and
therefore represent normalizable modes.

The basis states introduced above are complex. The physical deformations are obtained by
taking linear combinations of these states with complex coefficients so that the actual
deformation is real. Similar reality conditions must be imposed also on the fermionic
zero modes to be discussed below.

Since $E_0^\mu\xi_\mu$ has degree of growth 1, it would naively indicate that if we
apply this diffeomorphism on a (delta function) normalizable mode, we shall
produce a non-normalizable mode. However since $E_0$ acting on a mode
reduces the degree of growth by 1, the action of the diffeomorphisms generated by
\refb{ediffpar1} (and \refb{ediffpar2}) do
not increase the degree of growth of a mode and maps normalizable
modes to normalizable modes.

Let us now turn to the gravitino zero modes. We get from \refb{ed2bb}
\be\label{edefchiz}
\chi^\pm_{0,0}= e^{i\phi/2} \pmatrix{i\sin{\psi\over 2}\cr \pm \cos{\psi\over 2}},
\qquad
\eta^\pm_{0,0}=e^{-i\phi/2} \pmatrix{\cos{\psi\over 2}\cr \pm i\sin
{\psi\over 2}},
\ee
satisfying
\be
D_\alpha \chi^\pm_{00} =\pm \, {i\over 2}\, a^{-1} \, \gamma_\alpha \, \chi^\pm_{00} \, ,
\qquad D_\alpha \eta^\pm_{00} =\pm \, {i\over 2}\, a^{-1} \, 
\gamma_\alpha \, \eta^\pm_{00} , \qquad \alpha=\psi,\phi\, .
\ee
The gravitino zero modes take the form of pure gauge deformations as given in \refb{esusygrav}:
\be\label{egravitinozero}
\psi^{\pm s}_{k\mu} = \left( D_\mu \pm {i\over 2} 
a^{-1} \tau_3 \gamma_\mu\right) \eps^{\pm s}_k\, ,
\quad s=\pm, \quad k \in \ZZZ+{1\over 2}, \ k\ne \pm{1\over 2} \, ,
\ee
where, for $k>0$,
\ben \label{ekill1}
\eps^{\pm +}_k &=& \chi^\mp_{00} \otimes \left( \chi_{k-{1\over 2}}^+(i) + \chi_{k-{1\over 2}}^-(i)\right)
\mp i\, \chi^\pm_{00} \otimes \left( \chi_{k-{1\over 2}}^+(i) - \chi_{k-{1\over 2}}^-(i)\right)\, ,
\nonumber \\
\eps^{\pm -}_k &=& \eta^\mp_{00} \otimes \left( \chi_{k-{1\over 2}}^+(i) + \chi_{k-{1\over 2}}^-(i)\right)
\mp i\, \eta^\pm_{00} \otimes \left( \chi_{k-{1\over 2}}^+(i) - \chi_{k-{1\over 2}}^-(i)\right)\, ,
\een
and for $k<0$,
\ben\label{ekill2}
\eps^{\pm +}_{k} &=& \chi^\mp_{00} \otimes \left( \eta_{|{k+{1\over 2}}|}^+(i) + \eta_{|{k+{1\over 2}}|}^-(i)\right)
\mp i\, \chi^\pm_{00} \otimes \left( \eta_{|{k+{1\over 2}}|}^+(i) - \eta_{|{k+{1\over 2}}|}^-(i)\right)\, ,\nonumber \\
\eps^{\pm -}_{k} &=& \eta^\mp_{00} \otimes \left( \eta_{|{k+{1\over 2}}|}^+(i) + \eta_{|{k+{1\over 2}}|}^-(i)\right)
\mp i\, \eta^\pm_{00} \otimes \left( \eta_{|{k+{1\over 2}}|}^+(i) - \eta_{|{k+{1\over 2}}|}^-(i)\right)\, .
\een
Explicit form of $\eps^{\pm\alpha}_k$ can be found using \refb{ed2aa}, 
\refb{edefchiz} and the following expressions for the hypergeometric functions,
\be\label{ehyperspec}
F(m,m+2;m+1;z) = {m+1-z\over (m+1)(1-z)^{m+1}}, \qquad
F(m,m+2;m+2;z)={1\over (1-z)^m}\, .
\ee
In particular for $k=\pm{1\over 2}$, \refb{egravitinozero} vanishes and
\refb{ekill1}, \refb{ekill2} 
give us the Killing spinors \refb{ekillingspinor} up to normalization. Using  
\refb{ehypergrowth}, \refb{ed2aa} 
one can show that $\eps^{\pm\alpha}_k$ grow as $e^{\eta/2}$ for
large $\eta$ and are therefore not normalizable. However
one can show that the gravitino zero modes \refb{egravitinozero} 
are normalizable modes. A useful result for this proof is that the combinations
\be \label{eadd1a}
\left(D_b\pm {1\over 2a} \sigma_3
\gamma_b\right)\chi^\pm_m(i),
\qquad 
\left(D_b\pm {1\over 2a}  \sigma_3
\gamma_b\right)
\eta^\pm_m(i), \qquad \hbox{for} \quad m=1,\cdots \infty, \quad b=0,1 \, ,
\ee
are normalizable modes. This can be verified using the definitions of 
$\chi^\pm_m$ and $\eta^\pm_m$ given in \refb{ed2aa}.

The diffeomorphisms generated by the vector fields $\xi_\mu$ appearing in
\refb{ediffzero}, the supersymmetry transformations generated by 
\refb{ekill1}, \refb{ekill2}, the isometries of $AdS_2\times S^2$ and the Killing spinors
given in \refb{ekillingspinor} together are expected to generate an
$\NN=4$ super-Virasoro algebra, given by,\footnote{The generators given
here can be identified with the hatted generators given in \cite{0905.2686}.}
\ben\label{esp1}
&& \hskip -.7in [L_m, L_n] = (m-n) \, L_{m+n}\, , \nn
&& \hskip -.7in [J^3_m , J^\pm_n ] = \pm J^\pm_{m+n} \, , 
\qquad [J^+_m , J^-_n ] =  2\, J^3_{m+n}, \qquad J^\pm_n\equiv J^1_n\pm i J^2_n \, , \nn
&& \hskip -.7in [L_n, G^{\alpha\pm}_r] = \left({n\over 2}-r\right) 
G^{\alpha\pm}_{r+n}
\, , \nn
&& \hskip -.7in [J^3_m , G^{\alpha\pm}_r] = \pm\, {1\over 2}\, 
G^{\alpha\pm}_{r+m }\, , 
\quad 
[J^\pm_m , G^{\alpha\mp}_r] = G^{\alpha\pm}_{r+m}  \, ,\nn
&& \hskip -.7in \{G^{+\alpha}_r, G^{-\beta}_s\} = 2\, \ve^{\alpha\beta}\,
L_{r+s} - 2\, (r-s)\, 
(\ve\sigma^i)_{\beta\alpha}\, J^i_{r+s}\nonumber \\
&& \hskip -.7in \ve^{+-}=-\ve^{-+}=1, \quad \ve^{++}=\ve^{--}=0
\, , \qquad m,n=0,\pm 1, \quad r,s=\pm{1\over 2},
 \quad \alpha,\beta=\pm\, . 
\een
The generators $L_k$ are associated with the diffeomorphisms given in \refb{ediffpar1},
the generators $J^i_k$ are associated with the diffeomorphisms given in 
\refb{ediffpar2} and the generators
$G^{\alpha\beta}_k$ for $\alpha,\beta=\pm$
are associated with the supersymmetry transformations given in
\refb{ekill1}, \refb{ekill2}.
The subalgebra consisting of $L_0,L_{\pm1}$, $J^i_0$ and
$G^{\alpha\beta}_{\pm1/2}$ with $\alpha,\beta=\pm$ is expected to
generate the $PSU(1,1|2)$ group
of isometries of the $AdS_2\times S^2$ geometry. This includes the $SL(2,R)$ isometry
of $AdS_2$, the $SU(2)$ isometry of $S^2$ and the supersymmetries generated by the
Killing spinors $\eps^{\alpha\beta}_\gamma$ with $\alpha,\beta,\gamma=\pm$.
If $G$ denotes the group generated by the full superconformal algebra, then the coset
$G/PSU(1,1|2)$ describes the space of zero modes of the graviton and gravitino.
The action of the super-isometry group of $AdS_2\times S^2$ on the zero modes is given by
left multiplication of the coset element by an element of $PSU(1,1|2)$. 

If we define,
\be\label{edefQ}
Q=G^{++}_{-1/2}+G^{--}_{1/2}\, ,
\ee
then we have
\be
Q^2 = 2(L_0-J_0), \qquad J_0\equiv J^3_0\, .
\ee
These was used in \cite{0905.2686} to argue that the integration over the zero modes
can be localized into the $L_0-J_0$ invariant subspace of  $G/PSU(1,1|2)$.
This is the subspace spanned
by $J^+_{-1}$ and $J^-_{1}$ and is a finite volume subspace $SU(2)/U(1)=S^2$, yielding a
finite result from path integral over the zero modes.

However, given that the exponential growth of the Killing spinors invalidates the
proof of localization if we restrict the path integral to be over (delta-function) normalizable
modes, it is instructive to ask how the exponential growth of the Killing spinors
is compatible with the action of supersymmetry in the zero mode sector. Since 
 Killing spinors have degree of growth 1/2, one would expect that transformation by these
Killing spinors should increase the degree of growth of the modes
by 1/2, spoiling normalizability. 
This is the question we shall now address.

As a specific
example, consider the supersymmetry transformation of the gravitino zero modes. This is given
by the transformation generated by the anti-commutator of a supersymmetry transformation
generated by a Killing spinor given in \refb{ekillingspinor} with the 
supersymmetry transformation generated by $\eps^{\pm\alpha}_k$ given in \refb{ekill1},
\refb{ekill2} with $\alpha$ taking values $+$ or $-$.
Since both the Killing spinors and $\eps^{\pm\alpha}_k$ have degree of growth
$1/2$, the anti-commutator will be given by a transformation with degree of growth 1.
This seems to be in apparent 
conflict with the fact that for the vector fields $\xi_\mu$ in
\refb{ediffzero} that
generate the metric zero modes, only $E_0^\mu \xi_\mu$
has degree of growth 1 and $E_c^\mu \xi_\mu$ for other values of $c$ have
degree of growth $\leq 0$. We shall now show how this is achieved due to suitable cancellation.

We first note from \refb{ehypergrowth}, \refb{ed2aa}, \refb{edefchiz}, \refb{ekill1}, \refb{ekill2}
that  for large $\eta$, we have
\be\label{espinasy}
\eps^{\pm \alpha}_k \simeq e^{\eta/2} \pmatrix{v \cr \pm i\sigma_3 v}+\OO(e^{-\eta/2})\, , \quad \hbox{for
$\alpha=+,-$ and any $k$}\, ,
\ee
where $v$ is some two dimensional 
vector. Since the Killing spinors are given by $\eps^{\pm \alpha}_k$
with $k=\pm{1\over 2}$, they also have the same behaviour for large $\eta$. Now from
\cite{deWit:1979dzm} we know that the anti-commutator between the supersymmetry transformations
generated by $\eps^{\gamma\alpha}_k$ and $\eps^{\delta\beta}_l$ vanishes 
for $(\gamma,\delta)= (+,+)$ and $(-,-)$ and  the
anti-commutator between the supersymmetry transformations
generated by $\eps^{\pm\alpha}_k$ and $\eps^{\mp\beta}_l$,
acting on the
metric and the gravitino fields, produces
a diffeomorphism with
parameter proportional to
\be \label{edefxic}
\xi^c = \bar\eps^{\pm\alpha}_l \,  \gamma^c \eps^{\mp\beta}_k\, ,
\ee
where $\bar\eps$ has been defined in \refb{edefC}.
Since our interest is in studying how the unbroken supersymmetry acts on the zero modes,
we take $l=\pm{1\over 2}$. Using \refb{espinasy}, $\xi^c$ given in \refb{edefxic} may be expressed as,
\be
e^{\eta/2} \, \pmatrix{v^T & \pm i \, v^T\, \sigma_3}\, C\,  \gamma^c \,
e^{\eta/2}\, \pmatrix{u\cr \mp i\, \sigma_3 u} +\OO(1)
\ee
Now we have, in this notation,
\be
\gamma^0 = \pmatrix{ 0 & i\sigma_3\cr -i \sigma_3 & 0}, \quad \gamma^1 = 
\pmatrix{ 0 & \sigma_3\cr \sigma_3 & 0}, \quad \gamma_2 = 
\pmatrix{ -\sigma_2 &0\cr 0 & -\sigma_2}, \quad \gamma_3 = 
\pmatrix{ \sigma_1 &0\cr 0 & \sigma_1}\, .
\ee
This gives
\ben\label{ecancel}
&& \hskip -.3in e^\eta\, \pmatrix{v^T & \pm i \, v^T\, \sigma_3} \, C\,  
\gamma^0 \pmatrix{u\cr \mp i\, \sigma_3 u}
=2\, i\, e^\eta \, v^T \sigma_1 u, \quad 
e^\eta\, \pmatrix{v^T & \pm i \, v^T\, \sigma_3} \, C\,  
\gamma^1 \pmatrix{u\cr \mp i\, \sigma_3 u}
=\OO(1), \nonumber \\
&& \hskip -.3in  e^\eta\, \pmatrix{v^T & \pm i \, v^T\, \sigma_3} \, C\,  
\gamma^2 \pmatrix{u\cr \mp i\, \sigma_3 u}
=\OO(1), \quad 
e^\eta\, \pmatrix{v^T & \pm i \, v^T\, \sigma_3} \, C\,  
\gamma^3 \pmatrix{u\cr \mp i\, \sigma_3 u}
=\OO(1)\, . 
\een
This indeed agrees with the asymptotic form of the diffeomorphisms $\xi^\mu$
generating the metric
zero modes, as discussed below \refb{ediffzero}. In particular, note the cancellation of
the leading term in the second, third and fourth terms in \refb{ecancel} without which
we could not match the commutators of supersymmetry generators with the
asymptotic form of the diffeomorphism parameters $\xi$. Keeping track of the full
$\eta,\theta,\psi,\phi$ dependence one can also verify that the 
resulting diffeomorphism parameter
coincides with a linear combination of the parameters given in \refb{ediffpar1}, 
\refb{ediffpar2}  up to
diffeomorphisms by normalizable parameters. We shall describe a more general version of this
analysis below.

To complete the analysis we must also show that the action of the Killing spinor on a
metric zero modes produces a linear combination of 
the gravitino zero modes. In particular they must have the same degree of growth as
the gravitino zero modes given in \refb{egravitinozero}. For this we need to
compute the commutator between the supersymmetry transformation generated by the
Killing spinors and the diffeomorphism generated by the vector fields $\xi_\mu$ appearing
in \refb{ediffzero}. This can be identified as the action of the diffeomorphism generated
by $\xi_\mu$ on the Killing spinors. 
Since we have argued that the diffeomorphisms generated by the $\xi_\mu$'s do not increase
the degree of growth, the degree of growth of the commutator is the same as that of the
Killing spinors, i.e. 1/2. This agrees with the degree of growth of the supersymmetry
transformation parameters given in \refb{ekill1}, \refb{ekill2}, showing that the results are
consistent with the $\NN=4$ super-Virasoro algebra.

We can in fact verify the full $\NN=4$ super-Virasoro algebra given in \refb{esp1} using the 
supersymmetry algebra of $\NN=2$ supergravity theory. We shall illustrate this with one
example. According to \refb{esp1}, we have,
\be\label{e423}
\{G^{++}_k, G^{-+}_{k'}\}=-2\, (k-k') \, J^+_{k+k'}\, .
\ee
Let us for definiteness take $k,k'>0$.
Since the supersymmetry transformation parameters corresponding to $G^{++}_k$ and
$G^{-+}_{k'}$ are respectively $\eps^{++}_k$ and $\eps^{-+}_{k'}$ given in \refb{ekill1},
their anti-commutator will be diffeomorphism with parameter $\xi^\mu$, given by
\be
a\, E^b_\mu \xi^\mu = (\eps^{++}_k)^T C \gamma^b \eps^{-+}_{k'}\, .
\ee
Using \refb{ekill1}, \refb{ed2aa}, \refb{edefchiz}, \refb{ehyperspec},
\refb{edefgamma} and \refb{edefC} we get, up to an overall
constant of proportionality,
\be
a\, (E^0_\mu \xi^\mu,E^1_\mu\xi^\mu,E^2_\mu\xi^\mu,E^3_\mu\xi^\mu)= (k-k') \, 
e^{i(k+k')\theta+i\phi} \, {\sinh^{k+k'}\eta\over (1+\cosh\eta)^{k+k'}}\,
\left(0,0,-\cos\psi, i
\right)\, .
\ee
This can be written as,
\be
a\, (E^0_\mu \xi^\mu,E^1_\mu\xi^\mu,E^2_\mu\xi^\mu,E^3_\mu\xi^\mu)
= -2\, (k-k') \, {1\over 2}\, \Phi_{k+k'} \, \left( 0,0, \p_\psi Y_{11}(\psi,\phi),-{1\over \sin\psi}
\p_\phi Y_{11}(\psi,\phi)\right)\, ,
\ee
where $\Phi_k$ has been defined in \refb{e24pa}. Comparing this with \refb{ediffpar2} 
we see that
this is indeed the transformation parameters associated with the generator on the right
hand side of \refb{e423}. Similar analysis can be done to establish the other 
(anti-)commutators in \refb{esp1}, but in some cases 
we need to add to the final transformation
parameter generated this way  additional  general
coordinate / supersymmetry transformation
with normalizable parameters to get exactly the parameters that appear in
\refb{ediffpar1}, \refb{ediffpar2}, \refb{ekill1}, \refb{ekill2}. This is allowed
as the latter describe genuine gauge transformations.

From this analysis we see that supersymmetry transforms the zero modes of the metric and
graviton fields into each other and therefore we can use it to localize the path integral over
these zero modes, as anticipated in \cite{0905.2686}.
Ref.\cite{2209.13602} regulated the integration over these zero modes by turning on a small temperature that
destroys the isometry of the background geometry and produces an action for the zero modes
given by the super-Schwartzian. While this is a valid procedure and is needed for non-BPS
black holes, we see from our analysis that for BPS black holes, supersymmetric localization
can also yield a finite result for the integral over the zero modes.

We can contrast the situation for the zero modes
with the one where the Killing spinor is used to transform a
generic (delta-function) normalizable mode of the gravitino.
In particular, consider the transformation law of a vierbein in the presence of a
gravitino fluctuation. The transformation law is\cite{Ferrara:1976fu}
\be
\delta e^a_\mu \propto \bar\psi_\mu^{\pm} \gamma^a \eps^\mp
\ee
where $\eps^\mp$ is one of the Killing spinors. If we now pick 
the first basis state for the gravitino
described in \refb{egravitinobasis}, then the expression for
$\delta e^a_\mu$ described above will involve the combination 
$\chi^T \Gamma \eps^\pm$ for some combination $\Gamma$ of the $\gamma$-matrices.
However, unlike in the case of zero modes where the basis states have specific structure
as given in \refb{espinasy}, the basis states for the off-shell
delta-function normalizable modes of
a spinor
run over all four component spinors $\chi$ given by the product of \refb{ed2aa}
and \refb{ed2bb}
and therefore produces a result of order
\be
\delta e_\mu^a\sim e^{-\eta/2}e^{i\lambda\eta} e^{\eta/2} \sim e^{i\lambda\eta}\, .
\ee
This is not delta-function normalizable, showing that supersymmetry transformation of
a delta function normalizable mode takes us out of the space of delta-function normalizable 
modes.

Besides the zero modes of the metric and gravitino fields,
we also have zero modes of the gauge
fields. The associated field strengths vanish, and since supersymmetry transformation
involves the gauge fields only via their field strengths, the gauge field zero modes do not
participate in supersymmetry transformations. 
This can be traced to the fact that the commutator between two supersymmetry
transformations generates
a gauge transformation besides diffeomorphism\cite{deWit:1979dzm}, 
and if the gauge transformation parameter
is not normalizable then this gauge transformation cannot be removed. As a result
the algebra \refb{esp1} (or its $su(1,1|2)$
subalgebra) does not hold when acting on the gauge field
zero modes.
Therefore the path integral over the
gauge field zero modes cannot be localized. 
Apparently for each gauge field there are
infinite number of such zero modes, but by the procedure described in 
\cite{1005.3044,1106.0080,1108.3842}, this
infinity can be related to the infinite area of $AdS_2$. Since while computing the black
hole degeneracy we can throw away terms in the exponent proportional to the length of the
$AdS_2$ boundary\cite{0809.3304}, we can replace a term proportional to the
$AdS_2$ area in the exponent
of the partition function by the difference between the area and a multiple of
the boundary length. This
is a finite
number, and for each gauge field this gives the effective number of zero modes to be $-1$. 
For compact $U(1)$ gauge groups we expect the integration over these zero modes
to yield a finite result.

\sectiono{Logarithmic correction due to the zero modes} \label{slog}

In this section we shall compute the logarithmic correction due to the graviton and
gravitino zero modes
using localization results and compare this with the result of direct computation in
\cite{1106.0080,1108.3842}.

First let us review the direct computation.
One first normalizes the metric fluctuation $h_{\mu\nu}$ and gravitino fluctuation
$\psi_\mu$ by demanding the normalization condition on the modes:
\be
\int_{AdS_2\times S^2} \sqrt{\det g}\, g^{\mu\rho} g^{\nu\sigma} h_{\mu\nu} h_{\rho\sigma}
=1, \qquad \int_{AdS_2\times S^2} \sqrt{\det g}\, g^{\mu\nu} \bar\psi_\mu \psi_\nu =1\, .
\ee
Using the fact that $g_{\mu\nu}\propto a^2$, we get $h_{\mu\nu}\sim a^0$, $\psi_\mu\sim
a^{-1}$. Therefore we expand these fields as
\be\label{e51}
h_{\mu\nu} =\sum_b \phi_b \, f^b_{\mu\nu}, \qquad \psi_\mu = a^{-1} \sum_f \phi_f f^f_\mu\, ,
\ee
where $f^b_{\mu\nu}$ and $f^f_\mu$ are the zero mode wave-functions given as in 
\refb{ehmna}, \refb{e24pa}, \refb{egravitinozero}, 
normalized so that they do not carry any explicit
factors of $a$, and $\phi_b$ and $\phi_f$ are the bosonic and fermionic modes over
which we carry out the path integral.
On the other hands these modes can be represented as pure gauge
deformations:
\be\label{e52}
h_{\mu\nu}=D_\mu\xi_\nu+D_\nu\xi_\mu, \qquad \psi_\mu = (D_\mu+\cdots)\, \eps\, .
\ee 
where $\xi_\mu$ and $\eps$ are non-normalizable gauge transformation parameters.
We normalize $\xi^\mu$ and $\eps$ such that they directly generate
the transformations corresponding to the algebra \refb{esp1}  without any extra factor
of $a$. 
Since the action of the bosonic isometries on the coordinates
is $a$ independent ({\it e.g.} a shift in $\theta$ or $\phi$), 
one finds that we should take $\xi^\mu\sim 1$. Since $g_{\mu\nu}\sim a^2$, we have 
$\xi_\mu\sim a^2$.
Since supersymmetry
algebra relates $\bar\eps\gamma^\mu\eps$ to $\xi^\mu$, and since $\gamma^\mu\sim a^{-1}$,
one sees that $\eps\sim a^{1/2}$. Therefore we write
\be\label{e53}
\xi_\mu = a^2 \sum_b \chi_b \, g^b_\mu, \qquad \eps = a^{1/2} \sum_b \chi_f \, g^f\, ,
\ee
where $g^b_\mu$ and $g^f$ are the basis of non-normalizable deformations appearing in
\refb{ediffpar1}, \refb{ediffpar2},  \refb{ekill1}, \refb{ekill2} and $\chi_b$ and $\chi_f$
are the transformation parameters.  Using \refb{e51}, \refb{e52}
and \refb{e53} we now see that,
\be\label{e54}
\phi_b\sim a^2 \, \chi_b, \qquad \phi_f\sim a^{3/2}\, \chi_f\, .
\ee
Since integration over $\chi_b$ and $\chi_f$ generate the $a$-independent volume of the
group of superdiffeomorphisms of $S^1$, 
we see that integration over  each $h_{\mu\nu}$ zero mode 
integral produces a factor proportional to $a^2$ and  
the integration over each gravitino zero mode produces a factor of $a^{-3/2}$.

We now need to combine the above results with the zero mode counting results.
After subtracting the contributions proportional
to the length of the boundary of $AdS_2$, one finds that the number of zero modes
of the metric is $-6$ and the number of zero modes of the gravitino is 
$-8$\cite{1106.0080,1108.3842}. Since
each metric
zero mode integration gives a contribution of order $a^2$ and each gravitino zero mode
integration gives a contribution of order $a^{-3/2}$, the net $a$ dependence of
the partition function due to metric and gravitino zero modes is given by,
\be\label{e55}
a^{-6\times 2 + 8 \times 3/2} \sim a^0\, .
\ee
This is the essence of the analysis in \cite{1106.0080,1108.3842}
showing that the net contribution from integration
over the metric and gravitino zero modes is an $a$-independent result.

We shall now compare this with the result obtained from localization. First let us check that
the counting of the degrees of freedom after localization is consistent with the one before
localization. After localization we are left with no fermionic zero modes and
just two bosonic zero modes associated with
the generators $J^+_{-1}$ and $J^-_{1}$. 
Since there are no fermionic zero modes left,  the $-8$ fermionic zero modes that were
present before localization must
be paired with the $-8$ bosonic zero modes by the operator $Q$ defined in \refb{edefQ}.
Since we had $-6$ bosonic zero modes before localization, we should be left with
$8-6=2$ unpaired bosonic zero modes. These can be identified with the
zero modes associated with
the generators $J^+_{-1}$ and $J^-_{1}$.

Next we compute the contribution from the two bosonic zero 
modes that remain zero modes even after
switching on the deformation associated with localization.
Since these modes come from the metric, the integration over each 
zero mode gives a contribution $a^2$ according to \refb{e54}.
Therefore the net contribution from the zero modes
that are left after localization is
\be\label{elocz}
a^{2\times 2} = a^4\, .
\ee

We now need to combine this with the contribution from the $-8$
fermionic and bosonic modes
that are paired by $Q$ and hence cease to be zero modes once we switch on the deformation.
Naively one might think that their contributions will cancel, but this is not
the case. To illustrate this, let us suppose that we have a fermionic mode $\phi_f$ and
its bosonic partner $\phi_b$, each carrying non-zero $Q^2$ eigenvalue $h\sim 1$.
Since the algebra
\refb{esp1} does not have any explicit $a$ dependence, and since $Q$ is part of the algebra,
we have $Q\chi^b\sim \chi^f$ and
$Q\chi^f\sim\chi^b$. It then follows from \refb{e54} that
\be\label{e59}
Q\phi_b\sim a^{1/2} \phi_f, \qquad Q\phi_f \sim a^{-1/2} \phi_b\, .
\ee
Now suppose that we add a localizing term in the action proportional to
\be
Q(\phi_f Q\phi_f) \sim Q\phi_f Q\phi_f + \phi_f Q^2 \phi_f \sim a^{-1} \phi_b^2 + h\, \phi_f^2\, .
\ee
In this case the integration over each fermionic mode produces a factor of $a^0$, while
integration over each bosonic mode will produce a factor of $a^{1/2}$. The $-8$ bosonic modes
with non-zero $Q^2$ eigenvalue then gives a net factor of 
\be
a^{-8\times 1/2}\sim a^{-4}\, .
\ee
This cancels the $a^4$ factor in \refb{elocz}, giving a net contribution of order $a^0$. This
is in agreement with the result \refb{e55} that we find without using the localization
technique.

Note that \refb{e59} could also be derived without using the relation to the $\chi^f,\chi^b$
variables as follows. We simply use the fact that under supersymmetry\cite{Ferrara:1976fu} 
\be
\delta h_{\mu\nu}\sim \eps \, \gamma_{(\mu} \psi_{\nu)}\, .
\ee
$\gamma_\mu$ is of order $a$ since we need to contract $\gamma^b$ with $e_{b\mu}$.
On the other hand the Killing spinor $\eps$ carries a power of $a^{1/2}$ by the same 
argument that led to \refb{e53}. Using \refb{e51} we now get,
\be
\delta\phi_b \sim a^{1/2} \phi_f\, ,
\ee
in agreement with \refb{e59}. This shows that \refb{e59} holds  for  
general modes of the metric and
the gravitino and not just the modes that were zero modes in the undeformed theory.

\sectiono{Can we use localization to compute black hole entropy?} \label{sdiscuss}

Localization program for computing BPS black hole entropy, as envisaged in
\cite{0905.2686,1012.0265,1111.1161,1208.6221,1404.0033,1504.01400,1504.01700,
1806.03690,1806.04479,1904.08876,
2209.13602,2211.06873}, consists of two parts. First one localizes the path integral
to a finite dimensional subspace, known as the localization manifold,
by adding an appropriate term to the action that is
argued not to affect the result of the path integral. \refb{elocaction} represents such a term.
Second, one computes the integrand
that has to be integrated over the localization manifold. The result obtained this way
has close resemblance to the
Rademaker expansion of the microscopic formula for the black hole index, and the
challenge is to make this resemblance into an exact equality.

As we discussed in \S\ref{sconflict}, 
supersymmetry of the near horizon geometry, acting on delta-function normalizable
eigenmodes of the kinetic operator, takes us out of the space of such eigenmodes.
The path integral over these eigenmodes reproduces
correctly the
heat kernel\cite{campo,camhig1,campo2,camhig2,9505009} and  
the logarithmic corrections to the black hole entropy in a wide
variety of examples\cite{9709064,1005.3044,1106.0080,1108.3842,1109.3706,1404.1379},
and therefore seems to be the correct formulation of the path integral.
It would seem from these two observations that supersymmetry of the
near horizon geometry cannot be used to analyze the path integral over the near horizon
modes. This would imply that the first step of the localization program itself fails.
The exceptions are  the metric and gravitino zero modes, and localization yields a finite
result for integration over these modes. 

However, given that the localization 
program has the potential for bringing the gravitational path integral to the form of the
Rademaker expansion for the microscopic results on the black hole 
index\cite{0905.2686,1012.0265,1111.1161,1208.6221,1404.0033,1504.01400,1504.01700,
1806.03690,1806.04479,1904.08876,
2209.13602,2211.06873}, one might
wonder if there is an alternate formalism that could justify the application of
localization. Below we discuss some possibilities.
\begin{enumerate}
\item One possible option is to change the boundary conditions on the various fields in a way 
so that
supersymmetry does have well defined action
on the modes over which we integrate. This is the approach taken in
\cite{1806.04479}, and would correspond to
imposing normalizable boundary condition on the twisted fields in the language
of \cite{1802.00427}. Roughly, this corresponds to imposing standard boundary
condition on half of the fields and allowing non-normalizable fluctuations in certain
controlled way on the other half of the fields. However, this is
different from the standard boundary condition used in
\cite{campo,camhig1,campo2,camhig2,9505009} and also in the calculation
of logarithmic corrections to the black hole entropy 
in \cite{9709064,1005.3044,1106.0080,1108.3842,1109.3706,1404.1379}.\footnote{Note 
that while
\cite{9906056,1404.1379} used supersymmetry to organize the fields in $AdS_2$ after 
dimensional reduction on $S^2$, for the calculation of the logarithmic corrections
\cite{1404.1379} used the standard delta-function normalizable modes.}
Therefore  one needs to justify
the use of the new boundary conditions. In particular, a relevant question would be:
does the path integral over modes with standard boundary conditions yield the same
result as over the modes with supersymmetric boundary condition, not just at the
leading order but to all orders in the perturbation theory? 

It is possible that the
difference between the standard boundary conditions and supersymmetric boundary
conditions can be traced to the difference between the computation of degeneracy and
the computation of the index. Refs.\cite{1005.3044,1106.0080,1108.3842,1109.3706}
computed the degeneracy of states using gravitational path integral and there was a
separate argument showing the equality of degeneracy and index based on
supersymmetry\cite{0903.1477,1009.3226}. In contrast, 
\cite{2209.13602} attempts to compute the index directly using gravitational path integral.
If the difference in the choice of boundary condition comes from the difference between
index computation and degeneracy computation, 
one will need to justify why the
computation of the index requires this specific choice of boundary conditions. 
Some discussion on
boundary conditions in $AdS_2$ can be found in \cite{1802.00427} in the context of studying
quantum field theories on $AdS_2\times S^1$. Some of these issues have also been explored
in \cite{2302.10370}.
\item
If index computation requires the use of non-normalizable modes that
extend outside the near
horizon region, it seems appropriate to apply localization technique
on path integral over the modes
living in full asymptotically flat space-time geometry instead of
just on the modes in the near horizon geometry. Since Killing spinors approach constant
values at infinity in asymptotically flat space-time, we shall not encounter the kind of
issues we face by working  in the near horizon geometry.
Ref.~\cite{2011.06038}
developed a formalism
for computing the index of BPS black holes using path integral over supergravity fields in
the full asymptotically flat space-time geometry.
This could be a natural starting point for localizing the path integral. 
The added advantage of this approach will be that it will give the full
contribution to the entropy including any that may come from the hair modes living outside
the horizon, including the fermion zero modes associated with broken supersymmetry,
and will force us to directly compute the appropriate helicity weighted 
supertrace\cite{9611205,9708062} instead of the Witten index.
Furthermore in the general case, it will also get contribution from saddle points associated
with multi-centered black holes\cite{0010222}.
Since after localization we shall only pick saddle points that
are invariant under a combination of time translation and rotation about the z-axis, the
multi-centered configurations that would contribute are collinear arrangement of
black holes along the z-axis. This could make connection with the formula for the index
of multi-centered black holes in terms of single-centered black holes
given in \cite{1103.1887}, 
where also localization
and collinear configuration of black holes played a key role.
\item Zero modes represent on-shell field configurations.
The fact that localization can be used to perform the integration over the zero modes
suggests that the problem discussed here is related to off-shell field configurations
over which we carry out the path integral. Refs.~\cite{1802.00427,1906.02722}
discussed an alternate `on-shell' procedure for
computing derivatives of one loop determinants with respect to some parameters.  
A possible interpretation of this procedure is that we
carry out our analysis using a different basis of functions
where each basis state satisfies equation of motion as well as normalizable 
boundary conditions at large $\eta$ and smoothness at small $\eta$, but is allowed to 
violate equations of motion in the middle.
These basis states span the solution space of field equations with localized sources.
It was shown that
under favorable circumstances, this choice of basis
states transform into each other under supersymmetry generated by the Killing spinors.
If this holds for all the fields in the theory, and this type of basis states can be  used to 
compute the full path integral instead of just the derivative of the
one loop determinant, 
then one could in principle 
apply localization on the path integral over the coefficients of expansion
in this basis. It will be interesting to try to implement this on the supergravity
fields in the background of BPS black holes.
\end{enumerate}

Assuming that the first step in the localization program is successful, we are faced with
the second step: to what extent can we show that the integrand that appears in
the supergravity path integral after localization reproduces the terms in the
Rademaker expansion? The state of the art analysis on this can be found in
\cite{2209.13602} but as we discussed already,
there are several issues that need deeper investigation.
\begin{enumerate}
\item One needs to take into account the issues raised in \S\ref{snilp}. This could
require us to perform additional integrals over a finite set of bosonic and fermionic variables.
The other  possibility is that with another choice of boundary conditions compatible
with supersymmetry, the additional $L_0=J_0$ modes counted by the
index may go away. However this will require us to revisit the computation of the logarithmic
contribution to the entropy, since part of the logarithmic contributions comes from these
$L_0=J_0$ modes.
\item \cite{2209.13602} proposes a specific integration measure over the localization
manifold that has not been derived from first principles. Changing this will alter
the terms in the expansion, and the results would not agree with the Rademaker expansion.
Therefore it is important to find an {\it a priori} justification for the choice of measure
used in \cite{2209.13602}.
\item Finally, as discussed in \S\ref{suv}, the final expression for the black hole entropy, computed
via localization, is sensitive to the UV cut-off of the theory. Therefore a full understanding
of the problem will involve working with a UV complete description of the theory, like string theory.

\end{enumerate}

\bigskip

\noindent{\bf Acknowledgement:} I wish to thank Atish Dabholkar, Justin David, Edi Gava, 
Joao Gomes, Rajesh Gupta, 
Luca Iliesiu, Imtak Jeon, Sameer Murthy,
Kumar Narain and Gustavo Turiaci for useful
communications and patient explanation of their work.
This work was supported by ICTS-Infosys Madhava 
Chair Professorship and the J. C. Bose fellowship of the Department of Science
and Technology, India.

\end{document}